\documentclass[a4paper,12pt,oneside]{article}
\usepackage{fancyhdr}
\usepackage{pdfpages} 
\usepackage{mdwlist}
\usepackage{bm}
\usepackage{blindtext}
\usepackage{array}
\usepackage{multicol}
\usepackage{multirow}
\usepackage{mdwlist}
\usepackage{enumitem}
\usepackage[numbers]{natbib}
\usepackage{url} 
\usepackage{graphicx}
\usepackage{caption}    
\usepackage{amsmath}
\usepackage{amsthm}
\usepackage{amsfonts}
\usepackage{mathrsfs}
\usepackage{amssymb}
\usepackage{mdframed}
\usepackage{xcolor}
\usepackage{fourier-orns}
\usepackage{geometry}
\usepackage{parskip}
\usepackage{listings}
\usepackage{setspace}
\usepackage{imakeidx}
\usepackage{makeidx}
\usepackage{authblk} 
\usepackage{setspace} 
\usepackage{etoolbox}
\usepackage[
            pdftex
            ,breaklinks=true,
            ,plainpages=false
            ,pdfpagelabels
            ,pagebackref = false
            ,hyperindex
            ,breaklinks=true 
            ,colorlinks=true 
            ,citecolor=black 
            ,linkcolor=black
            ,urlcolor=black
            ,filecolor=black
            ,bookmarksopen=true
            ]{hyperref}

\def\be{\begin{equation}}
\def\ee{\end{equation}}

\makeindex

\title{\Large Roullete curves, coin paradox and Aristotle's wheel paradox}
\author[1]{Osvaldo L. Santos-Pereira}
\affil[1]{Universidade Federal do Rio de Janeiro, Rio de Janeiro, RJ, Brasil.}

\date{\today}

\begin{document}
\maketitle

\begin{abstract}
This work discusses the concept of roulette, the generated curves that occur when one curve rolls without slipping along another, tracing the path of a fixed point. The coin paradox and Aristotle's wheel paradox are used as pedagogical motivations to discuss the parametric equations of epicycloids and hypocycloids, providing a geometrical intuition for the mathematical derivations and computational implementation of those curves. Python code is provided to motivate the application of the derived parametric equations, resulting in concrete visualizations and animations.
\end{abstract}

\textbf{Keywords:} Roullete; Aristotle; paradox; coin; wheel; epicycloid; hypocycloid

\section{Introduction}

Roulettes are generated when one curve rolls without slipping along another, tracing the path of a fixed point. Classic examples include cycloids, epicycloids, hypocycloids, and cardioids \cite{Maxwell1849, Lockwood1967, Cundy1989, Gardner1984, Yates1952, Lawrence1972, Zwillinger1996}. Despite their simple construction, roulette often yields surprising and counterintuitive results. 

Two famous cases illustrate how misconceptions can arise from the analysis of roulette curves, examples are the coin paradox and the Aristotle wheel paradox. An epicycloid consists of an exterior circle rolling about another fixed circle, and the coin paradox was first defined as a particular case of an epicycloid: a coin rolling about another coin of equal radius completing two full turns, not one, as intuition suggests \cite{Gardner1975, Pappas1989, Steinhaus1999}. In Aristotle’s wheel paradox, concentric wheels of different radii seem to traverse the same distance, apparently contradicting the relationship between circumference and radius \cite{Drabkin1950, Ballew1972, Costabel1968}.

The coin paradox has an analogue in celestial mechanics. The concept of sidereal time. A sidereal day \cite{Bartlett1904} represents the time for one rotation about the planet's axis relative to the distant stars, just like the rolling coin in the paradox. An Earth year has about 365.25 solar days but about 366.25 sidereal days. As a solar day has 24 hours, a sidereal day has approximately 23 hours, 56 minutes, and 4.1 seconds. 

The coin paradox is often illustrated through the epicycloid curve, traced by a rolling coin on a fixed coin of the same size. An epicycloid is a geometric curve traced by a point on a circle, an epicycle, that rolls without slipping around the outside of a fixed circle. The curve features points where it comes into contact with the fixed circle, and the number of cusps $N$ is equal to the ratio of the radius $R$ of the fixed circle to the radius of the rolling circle $r$, so $N = R/r$ when this ratio is an integer. 

Abad \cite{Abad2021} presented an interesting application of roulette curves, where he calculated the magnetic field generated by a steady current that takes the shape of two types of special curves: hypocycloids and epicycloids with a specified number of sides. The computation was performed at the center of the curves referred to. He used the Biot-Savart law, and the result was shown to be general because it depends on the number of sides of the curve and on a parameter that identifies the type of curve considered.

This paper provides a pedagogical introduction to roulette curves, highlighting their mathematical formulation, physical interpretation, and educational potential. Special attention is given to epicycloids, hypocycloids, and the paradoxes of rolling motion: the coin paradox and the Aristotle wheel paradox. One of the intentions of this paper is to demonstrate how these paradoxes connect with the intuitive aspects of geometry and mechanics in the context of physics teaching. 

This work is organized as follows: the second section presents and discusses the coin paradox, solving it using simple circular kinematics, which can be learned from fundamental physics textbooks \cite{Halliday2013}. The third section presents the epicycloid curves, derives the parametric equations for the curves, and provides several examples. The fourth section introduces the hypocycloid, where a simple change of signs in $r$ is made as an analogy to the derivation of the parametric equations for the epicycloids. Several curves are shown for different values of $k$. Section five presents and explains Aristotle's wheel paradox in the context of fundamental physics \cite{Halliday2013}. Section six presents all the Python codes used in this work to generate images and animations. The last section is the conclusion of this paper.

\section{Coin paradox} \label{coinpar}

The coin paradox was initially proposed by Martin Gardner in 1975 \cite{Gardner1975}. Gardner was a mathematician known for writing recreational mathematical puzzles. This apparent paradox gained notoriety in 1982 when it was posed as a SAT question. The SAT is a standardized test widely used for college admissions in the United States. The problem was that there was no correct answer among the five alternatives. Someone noticed that the question could not be answered correctly and reported the error to the company that administered the SAT, prompting a backlash widely reported in a respected U.S. newspaper \cite{nytimes}. 

The coin paradox was again brought up in 2015 on the streaming platform YouTube by a channel named \textit{MindYourDecisions}, in a video titled ``Why did everyone miss this SAT Math question?'' \cite{MYD} uploaded in july 5th of 2015. As of October 2025, the video has more than three million views.

An article named "The SAT Problem That Everybody Got Wrong" was written by Jack Murtagh and edited by Jeanna Bryner and published in an article in the Scientific American Journal in June 20 of 2023 \cite{sciam}. The article was about the 1982 SAT problem and the coin paradox.

On November 30th, 2023, the YouTube channel \textit{Veritassium} also uploaded a video regarding the coin paradox and the SAT question. As of October 2025, the Veritassium video named ``The SAT Question Everyone Got Wrong'' \cite{veritsat} has over sixteen million views. A fun fact: Dr. Doug Jungreis, now a researcher in mathematics and one of the people who reported the error in the 1982 SAT question, appeared in the video and was interviewed by Dr. Derek Muller, the owner of the Veritasium channel.

The coin paradox is as follows: start with two identical coins on a table, both showing their heads side up and aligned. Keep coin B fixed, and roll coin A around it without slipping. When coin A reaches the opposite side, the heads are again parallel, and coin A has completed one full revolution. Continuing the motion back to the starting point adds a second revolution. Paradoxically, coin A seems to have rolled a distance twice its circumference. 

Fig.~\ref{fig1} below illustrates one of the reasons why this paradox seems to appear in people's reasoning. The image shows five snapshots of the rolling coin on a fixed coin: the first one in the starting position, with its head in an upright position; the second snapshot, after the rolling coin has turned its head upside down; the third snapshot shows the rolling coin after completing half of the circular path with its head in an upright position again, the fourth snapshot shows the coin again with its head upside down but on the left of the fixed coin, and the final snapshot where the rolling coin returns to its starting point. In general, people will count the number of times the top of the rolling coin touches the fixed coin, imagining that this is half of the total roll.

\begin{figure}[ht]
    \centering
    \includegraphics[width=0.50\textwidth]{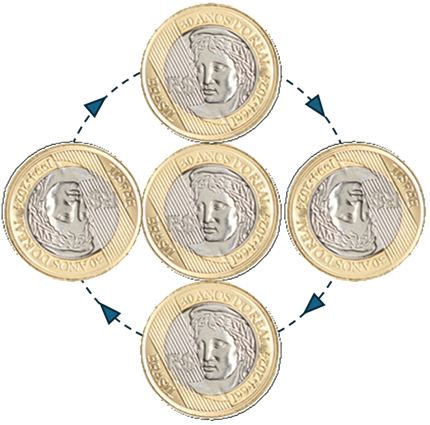}
    \caption{Illustration of the coin rotation paradox: a coin rolling externally around another identical coin completes two full rotations before returning to its initial position. The image shows a Brazilian one real coin (1 BRL).}
    \label{fig1}
\end{figure}

The explanation is that coin A’s circumference equals the distance rolled, but an additional rotation arises because its path is circular rather than straight. Flattening coin B’s circumference into a line shows only one actual rotation, while the second corresponds to the transport of A around B.

The center of the rolling coin follows a circular path of radius equal to the sum of both radii, so its trajectory has twice the circumference of either coin. Covering this distance without slipping forces the rolling coin to complete two full revolutions. Details of the coin’s orientation or rotation direction do not affect this result.

\begin{figure}[ht]
    \centering
    \includegraphics[width=0.45\textwidth]{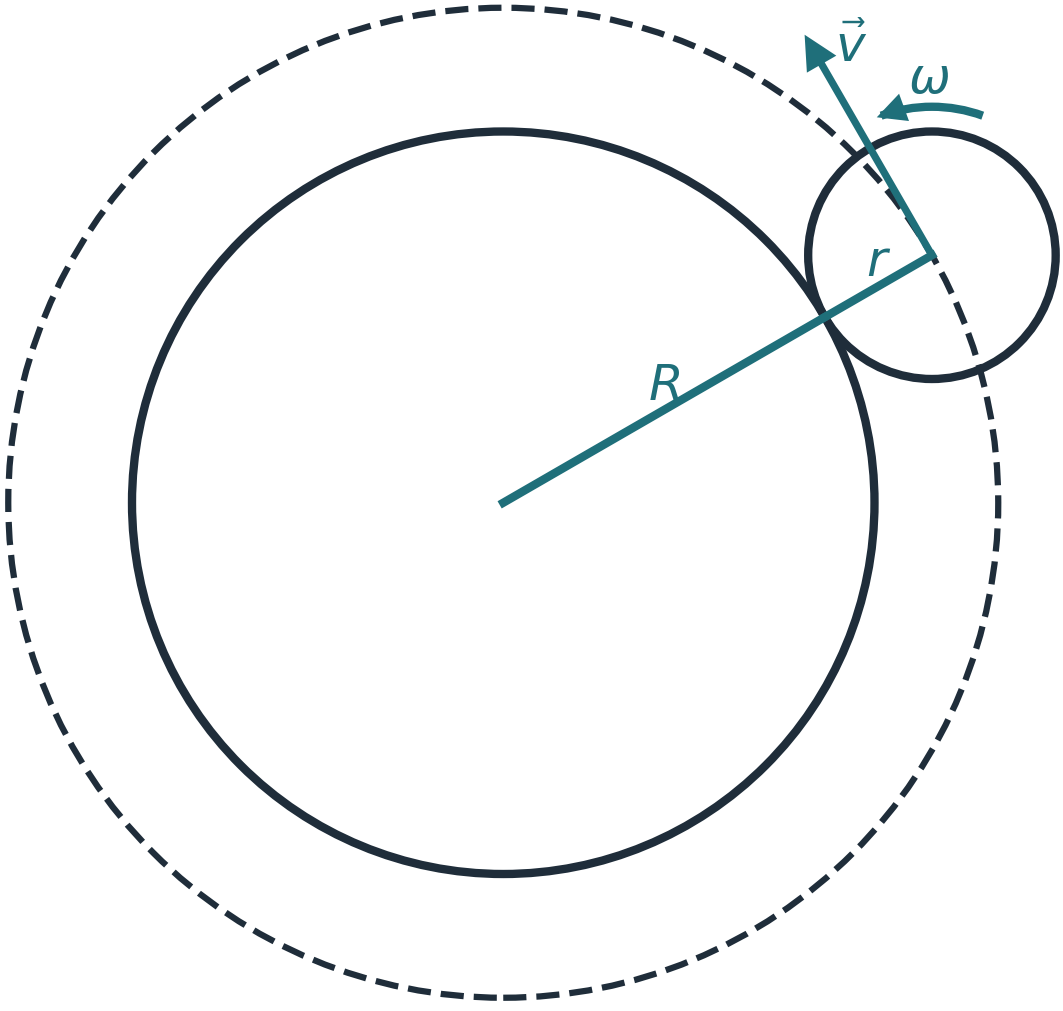}
    \caption{Geometry of a circle of radius $r$ rolling without slipping around a fixed circle of radius $R$. The center of the moving circle traces a circular path of radius $R+r$, leading to $(R+r)/r$ full rotations about its own axis.}
    \label{fig2}
\end{figure}

This paradox is easily explained using simple kinematics of circular motion. Consider a circle $A$ of radius $r$ rolling without slipping on the outside of a fixed circle $B$ of radius $R$, as shown in Fig.~\ref{fig2}, the center of circle $A$ traces a circular path of radius $R+r$.  

Let the linear speed of the center of $A$ be $v$, and its angular speed about its own center be $\omega$. The no-slipping condition at the point of contact between the two circles implies that the linear speed of the circle A's center is 
\begin{equation}
    v = \omega r.
\end{equation}
The time taken for the center of $A$ to complete one full revolution around $B$ is
\begin{equation}
t = \frac{2\pi (R+r)}{v} = \frac{2\pi}{\omega}\left(\frac{R}{r} + 1\right).
\end{equation}
During this interval, the angular distance $\theta = \omega t$ traversed by circle $A$ about its own center is given by the following expression
\begin{equation}
    \theta = 2\pi \left(\frac{R}{r} + 1\right).
\end{equation}
Hence, the number of complete rotations of $A$ about its own axis is
\begin{equation}
    N = \frac{\theta}{2\pi} = \frac{R}{r} + 1.
\end{equation}
As a particular case of the coin paradox, if $R = r$, then
\begin{equation}
N = \frac{r}{r} + 1 = 2,
\end{equation}
meaning that circle $A$ completes two full rotations about its center while rolling once around circle $B$.  

\section{Epicycloid}

\begin{figure}[h]
    \centering
    \includegraphics[width=0.5\textwidth]{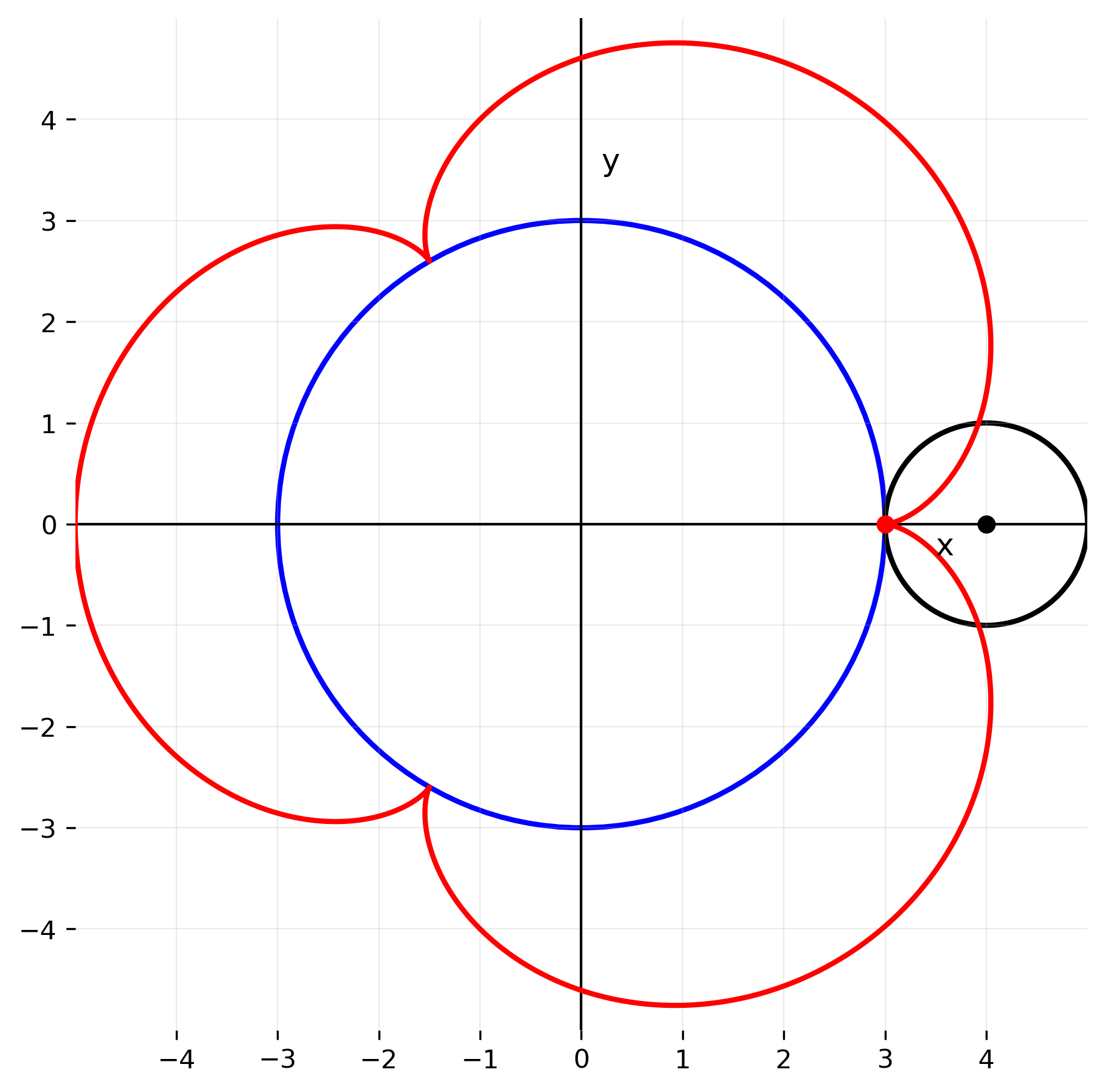}
    \caption{Example of an epicycloid generated by a circle of radius $r$ rolling around a fixed circle of radius $R$. The number of cusps depends on the ratio $R/r$. In this figure, $R = 3r$, and the number of turns completed by the smaller circle equals 4. The name of this epicycloid is trefoiloid \cite{Lawrence1972}.}
    \label{fig3}
\end{figure}

An epicycloid is a type of roulette curve generated by a fixed point on the circumference of a circle of radius $r$ that rolls without slipping around the outside of a fixed circle of radius $R$, just like the example we explored in Sec.~\ref{coinpar} about the coin paradox. The locus of the point traces a curve with a wide variety of shapes depending on the ratio $k = R/r$. For $k=3$ as shown in Fig.~\ref{fig3}, there is a trefoiloid \cite{Lawrence1972}. In this example, it would be the case of a circle of radius $r$ rolling without slipping around a circle B of radius $R=3r$.

Epicycloids are encountered in various physical and engineering contexts. They describe the motion of planetary gears \cite{Uicker2003, Paul1979} and caustic curves in optics \cite{Boyle2014}, with applications in kinematics and computer-aided design. They also provide elegant didactic examples of how rolling motion translates into complex yet tractable geometrical trajectories.

\begin{figure}[ht]
    \centering
    \includegraphics[width=0.7\textwidth]{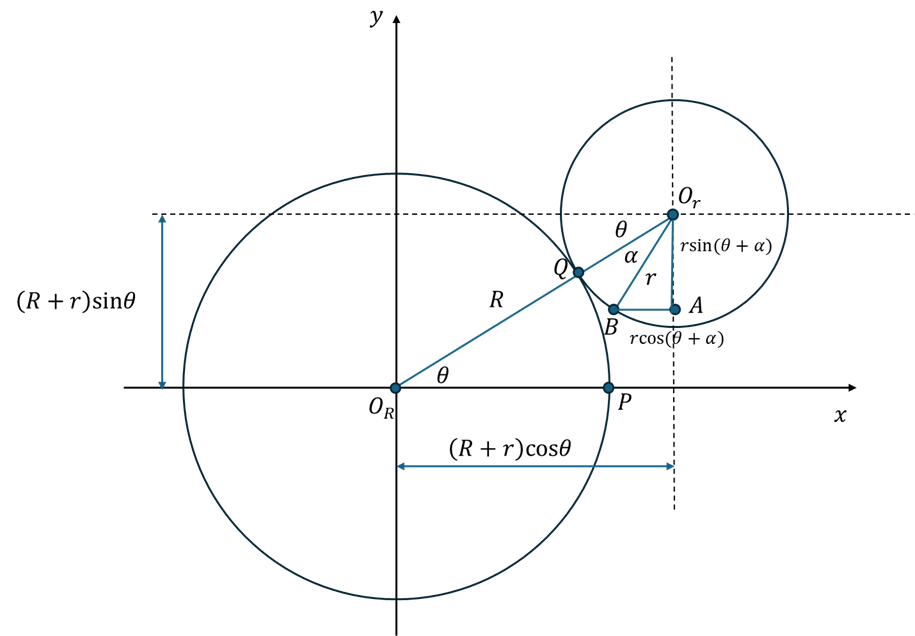}
    \caption{Construction of the epicycloid. A generating circle of radius $r$ rolls externally around a fixed circle of radius $R$, and a point $B$ on its circumference traces the epicycloid curve.}
    \label{fig4}
\end{figure}

It is straightforward to demonstrate the parametric equations of the epicycloid by analyzing the geometry of rolling without slipping. Consider a circle of radius $r$ rolling externally without slipping on a fixed circle of radius $R$, as illustrated in Fig.~\ref{fig4}. Consider the origin of the coordinate system at the center of the larger circle, at $O_R(0,0$). A point $B = (x,y)$ on the circumference of the rolling circle traces the epicycloid. 

Ref.~\cite{Lawrence1972} provides a visual proof of the parametric equations of a general roulette (\textit{epitrochoid}), traced by a point P attached to a circle that is rolling about a fixed circle. See Figs. 56-58 in chapter 6 of Ref.~\cite{Lawrence1972}. Fig.~\ref{fig4} was based on the derivation provided by Ref.~\cite{Lawrence1972}, however the reader can find several other proofs on the internet, in personal webpage of Professors, public archives or free online encyclopedias such as Wikipedia. It is customary for calculus and analytical geometry textbooks to leave this proof as an exercise for students. It is a relatively trivial exercise, but it must be stated that Fig.~\ref{fig4} is not an original proof, since the study of roulette dates from 1525 (D{\"u}rer) \cite{Lawrence1972}.

Let $\theta$ denote the angle corresponding to the arc length on the fixed circle measured from the starting tangent point of point $B_0 = (x_0,y_0)$, and let $\alpha$ be the angle corresponding to the arc length on the rolling circle measured from the contact point to the point $B$. Notice that at $\theta = 0$ the initial point is at $B_0 = (R,0)$.

The coordinates of the point $B$ on the rolling circle can be derived from geometric and trigonometric relations shown in Fig.~\ref{fig4}.
\begin{align}
x(\theta) &= (R+r)\cos\theta - r\cos(\theta + \alpha), \label{xcoord} \\
y(\theta) &= (R+r)\sin\theta - r\sin(\theta + \alpha). \label{ycoord}
\end{align}
Since the rolling occurs without slipping, the two arc lengths of the fixed circle $\widehat{PQ} = \theta R$ and the arc length of the rolling circle $\widehat{BQ} = \alpha r$ must be equal. Thus, the no-slipping condition gives the following relation
\begin{equation}
\alpha = \frac{R}{r}\,\theta. \label{alpha}
\end{equation}
Substituting Equation \eqref{alpha} in Equations \eqref{xcoord} and \eqref{ycoord} results in the following formulas
\begin{align}
x(\theta) &= (R+r)\cos\theta - r\cos\left(\frac{R+r}{r}\theta\right), \label{xcoordfinal} \\
y(\theta) &= (R+r)\sin\theta - r\sin\left(\frac{R+r}{r}\theta\right). \label{ycoordfinal}
\end{align}
These are the parametric equations of the epicycloid, expressed in terms of the rolling angle $\theta$. The structure of the curve, including the number of cusps, depends directly on the ratio $k = R/r$. 

The ratio $(R+r)\theta/r$ can be physically interpreted as the relative angular speed of the rolling circle compared to its translational motion around the fixed circle. For every increase of $2\pi$ in $\theta$, one complete revolution of the center around the fixed circle, the rolling circle rotates through an angle of $2\pi (R+r)/r$. Thus, the rolling circle completes not one, but $(R+r)/r$ complete rotations about its axis while making a single revolution around the fixed circle. This is a source of the paradox observed with the rolling coin, since the rolling motion and the transport of the circle’s center is the combined motion of a circle of radius $R+r$ and not only $r$, producing more rotations than intuition alone might suggest.

\begin{figure}[h]
    \centering
    \includegraphics[width=0.9\textwidth]{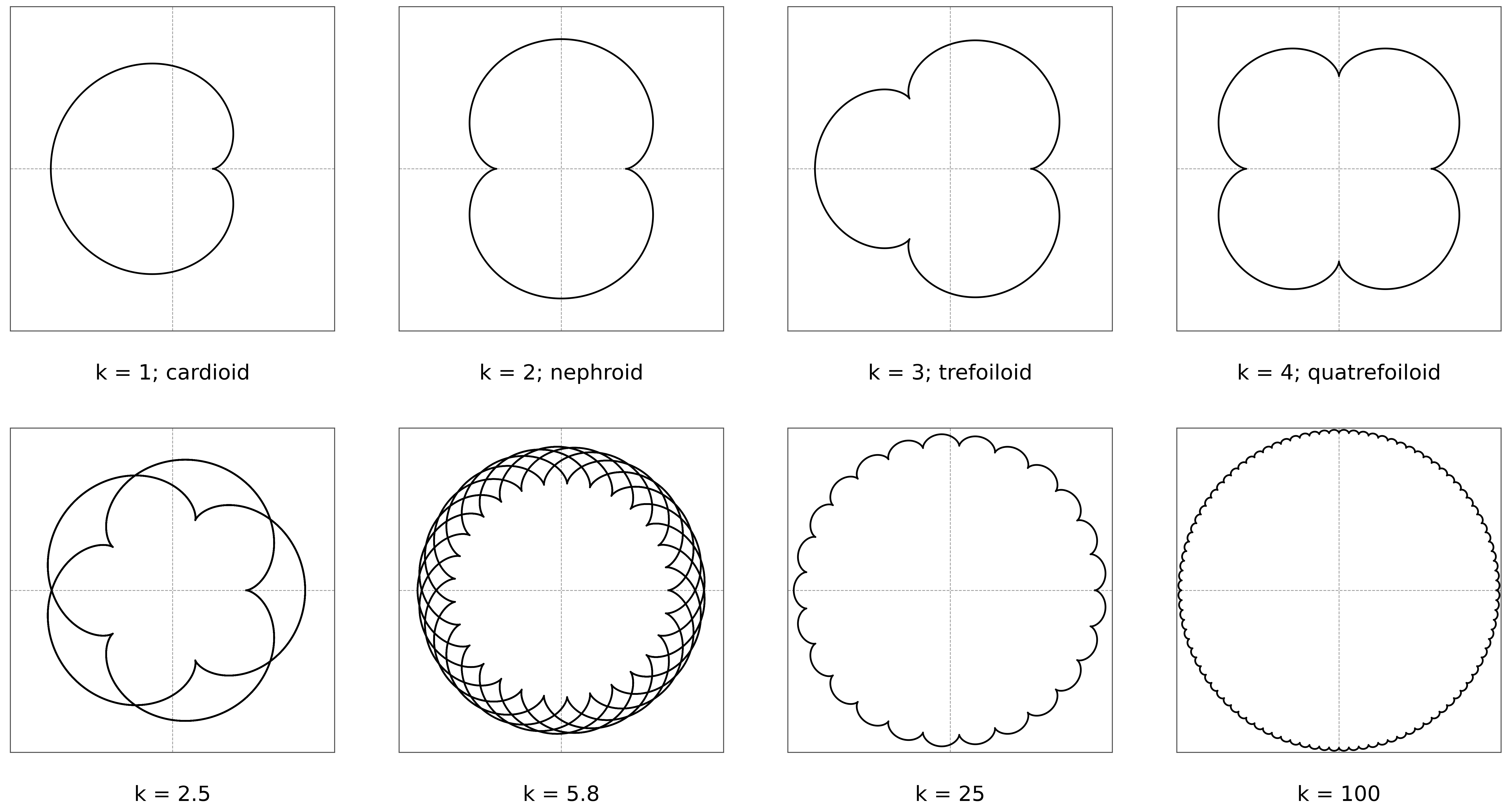}
    \caption{Examples of epicycloids for different values of $k = R/r$.}
    \label{fig5}
\end{figure}

Figure~\ref{fig5} illustrates several examples of epicycloids for different values of the ratio $k = R/r$, where $R$ is the radius of the fixed circle and $r$ is the radius of the rolling circle. When $k$ is an integer, the epicycloid closes with $k+1$ cusps, producing well-known curves such as the cardioid ($k=1$), the nephroid ($k=2$), the trefoiloid ($k=3$), and the quatrefoiloid ($k=4$). When $k$ is non-integer or rational, the curve is more intricate and may not close after a single revolution, generating dense, lace-like structures. 

\begin{figure}[h]
    \centering
    \includegraphics[width=0.6\textwidth]{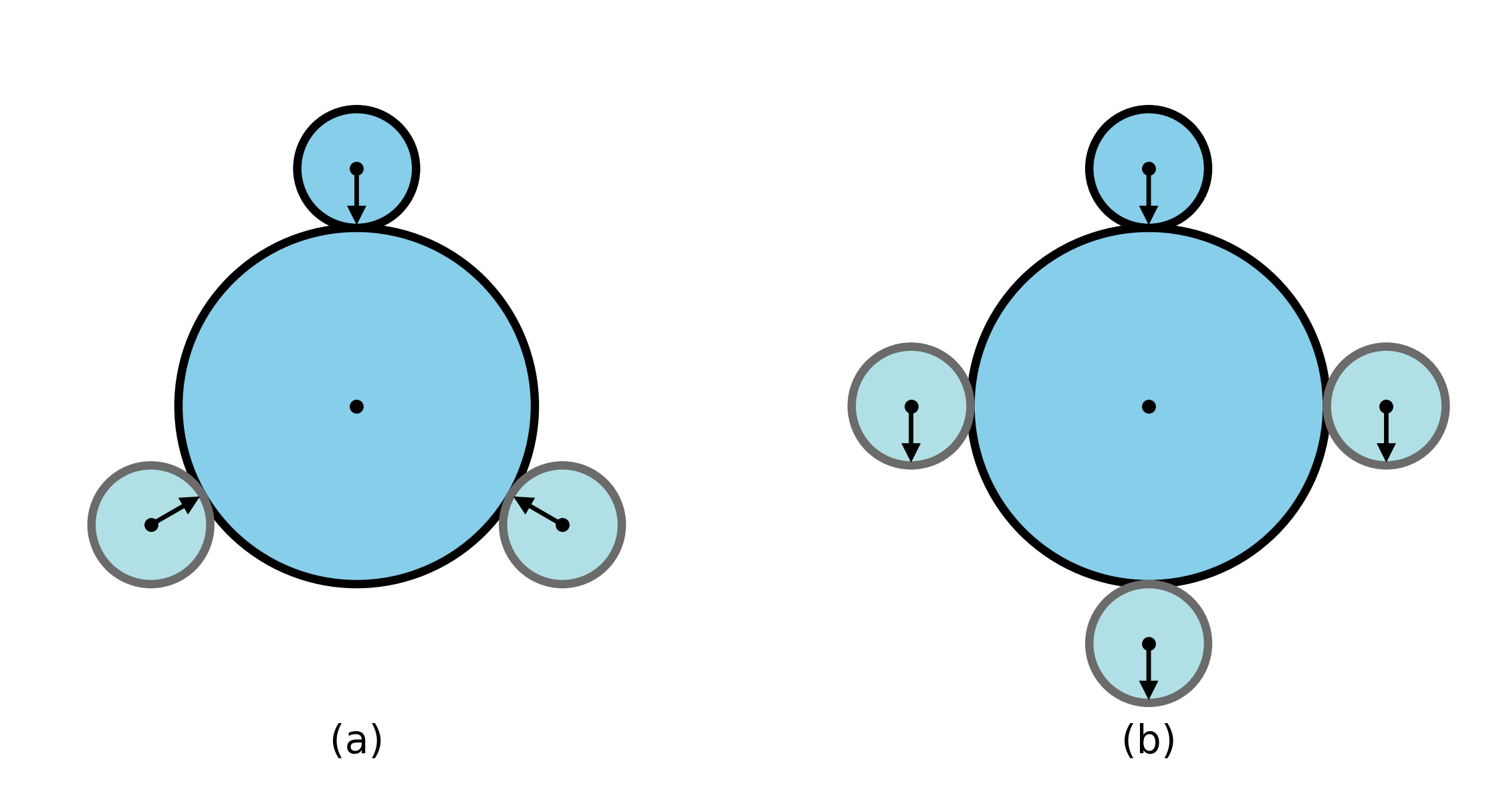}
    \caption{Illustration of the rolling paradox. 
    \textbf{Left:} the intuitive but incorrect reasoning, where the small circle is assumed to rotate only $R/r$ times while rolling around the larger circle. 
    \textbf{Right:} the correct reasoning: the circular path of the rolling circle’s center adds one extra rotation, giving $N = (R+r)/r$.}
    \label{fig6}
\end{figure}

Fig.~\ref{fig6} shows two points of view for the coin paradox. Fig.~(a) on the left indicates that the small coin is drawn with an arrow that initially points towards the center of the large coin. From the point of view of the large coin, the small coin will have completed a full rotation when the arrow once again points to the center of the large coin. This happens after 120 degrees, then again after 240 degrees, and finally after a full rotation of 360 degrees. From this viewpoint, the small coin rotates three times, and each rotation corresponds to the coin rotating along one-third of the large coin's circumference. So there is no paradox here. However, when we view the situation from an external perspective, we see an additional rotation, because while rolling along the edge of the large coin, the small coin also moves in a large circle. Fig.~(b) on the right shows that the small coin actually rotates 4 times before it returns to its initial position. This might seem very strange. If the bigger coin has 3 times the circumference of the small coin, and the small coin rotates without slipping, how can it possibly take four rotations to get back to its starting point? But it really does. 

\section{Hypocycloid}

A hypocycloid is generated by the trace of a fixed point on a small circle that rolls within a larger circle \cite{Lawrence1972}. This would be analogous to a coin rolling on the inside of a loop. As the radius of the larger circle is increased, the hypocycloid becomes more like the cycloid, which is the curve traced by a point on a circle as it rolls along a straight line without slipping. A cycloid is a specific form of trochoid. Fig.~\ref{fig7} below illustrates one example of a hypocycloid for $k=3$, named deltoid \cite{Lawrence1972}.

\begin{figure}[ht]
    \centering
    \includegraphics[width=0.5\textwidth]{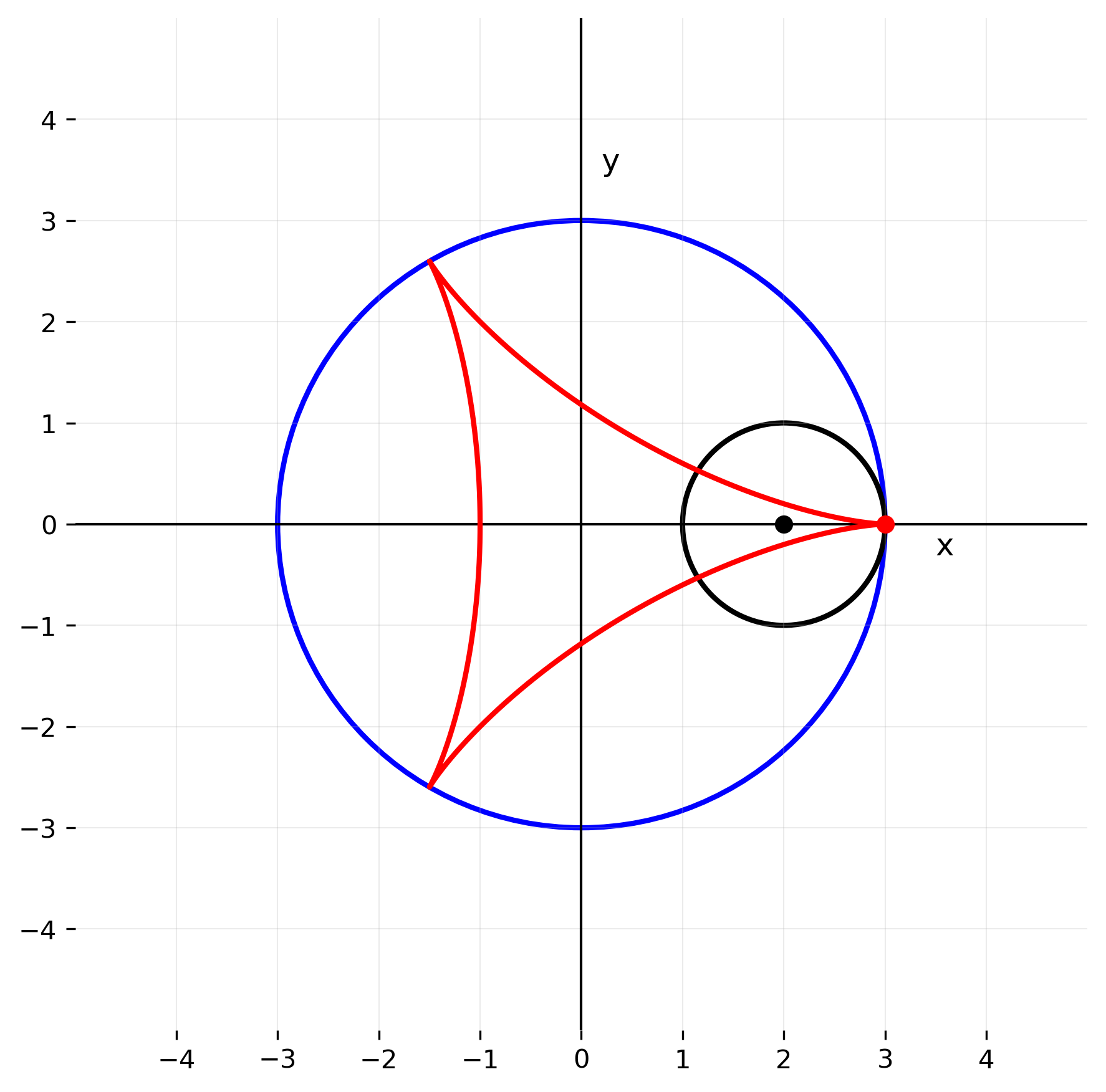}
    \caption{Examples of hypocycloids for $k = 3$, named deltoid \cite{Lawrence1972}.}
    \label{fig7}
\end{figure}

The derivation of the parametric equations for the hypocycloid is analogous to the derivation of the Epicycloid, using a similar figure as the one in Fig.~\ref{fig4}, but there is a simple way to get the equations immediately: use the change $r \to - r$ instead to get the following expressions
\begin{align}
x(\theta) &= (R-r)\cos\theta + r\cos\left(\frac{R-r}{r}\theta\right), \label{xcoordfinalhyp} \\
y(\theta) &= (R-r)\sin\theta - r\sin\left(\frac{R-r}{r}\theta\right). \label{ycoordfinalhyp}
\end{align}
Now the number of complete turns is diminished by one instead of the addition of a turn, as the rolling circle completes the $2\pi$ cycle. One interesting property of a hypocycloid is that for any value of r, it is a brachistochrone for the gravitational potential inside a homogeneous sphere of radius R, see pp. 230–2 of Ref.~\cite{Rana2001}.

\begin{figure}[ht]
    \centering
    \includegraphics[width=0.9\textwidth]{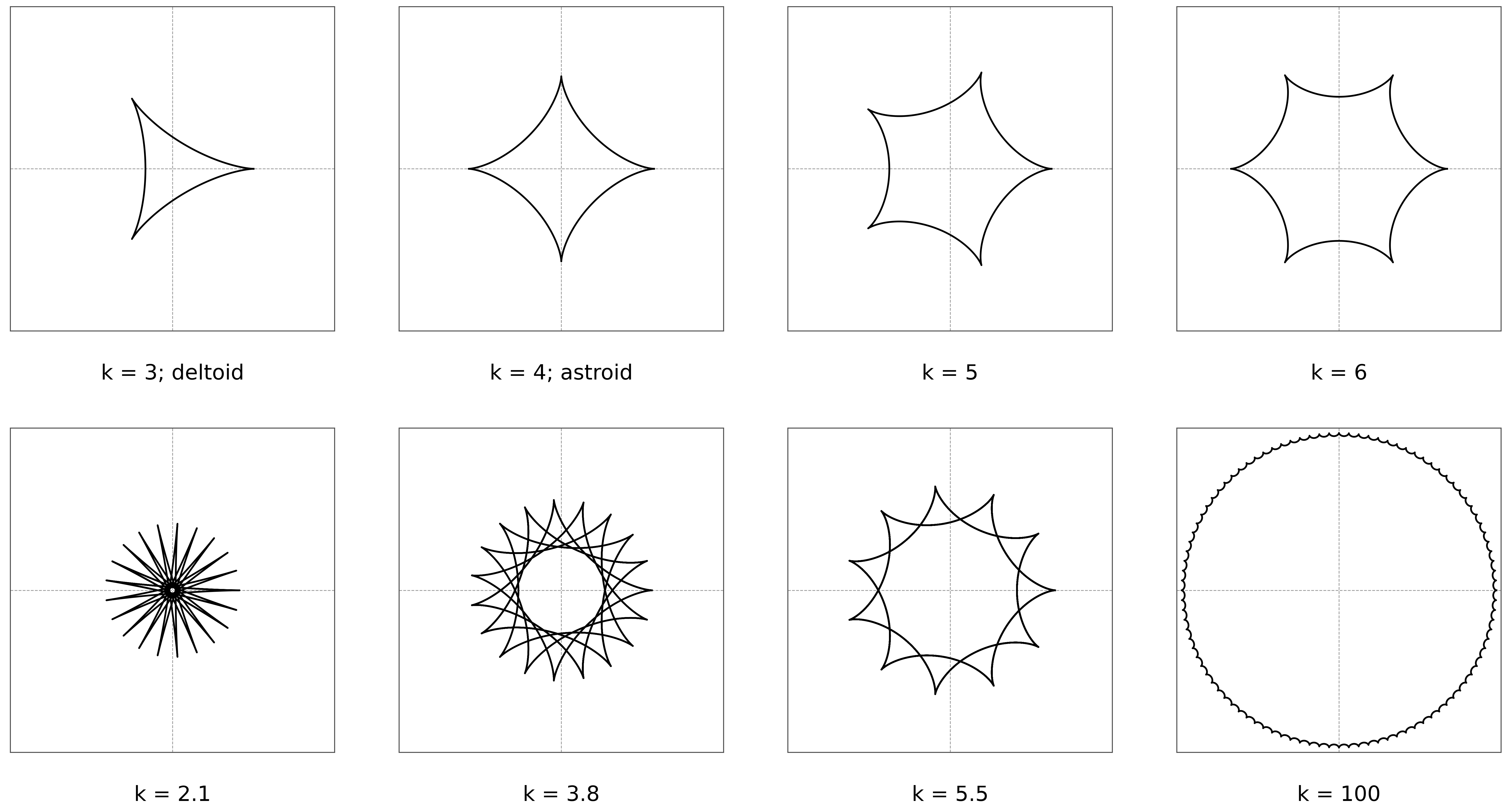}
    \caption{Examples of hypocycloids for different ratios $k = R/r$.}
    \label{fig8}
\end{figure}

As illustrated in Fig.~\ref{fig8}, hypocycloids exhibit a wide variety of shapes depending on the ratio $k = R/r$. When $k$ is an integer, the curve is closed and presents exactly $k$ cusps: for $k=3$, it gives the deltoid, $k=4$, the astroid, and larger integers produce more complex star-shaped forms. If $k$ is not an integer, the curve no longer closes after a single revolution of the rolling circle. Instead, the trajectory densely fills a region with intricate star-like patterns. This sensitivity to the value of $k$ highlights the richness of hypocycloids as pedagogical examples, providing a natural way to connect geometry, periodicity, and the role of rational versus irrational ratios in mathematics. 

\section{Aristotle's wheel paradox}

A classic example of a geometrical paradox is known as \emph{Aristotle's wheel paradox}. The setup consists of two concentric circles of different radii, rigidly connected, as if they were part of the same wheel. The larger circle rolls without slipping along a straight line.  

Intuitively, one might expect that if the larger circle of radius $R$ covers a distance equal to its circumference $2\pi R$, then the smaller circle of radius $r<R$ should simultaneously cover a distance equal to its own circumference $2\pi r$. This reasoning seems natural from the perspective of the individual kinematics of each circle. However, since the smaller circle is rigidly attached and rotates together with the larger circle, both circumferences appear to travel the same linear distance, despite their different radii and circumferences, leading to the paradoxical conclusion that arcs of various sizes are equal in length.  

The resolution of the paradox lies in observing the path traced by the smaller circle. Although it is concentric with the larger one, the smaller circle does not \emph{roll} on the ground. Instead, each point on its circumference describes a curve that is no longer a straight line but a \emph{roulette}. Consequently, the smaller circle does not actually unroll its full circumference along the ground. The apparent contradiction arises only if one assumes, incorrectly, that both circles roll in the same manner.  

\begin{figure}[ht]
    \centering
    \includegraphics[width=0.9\textwidth]{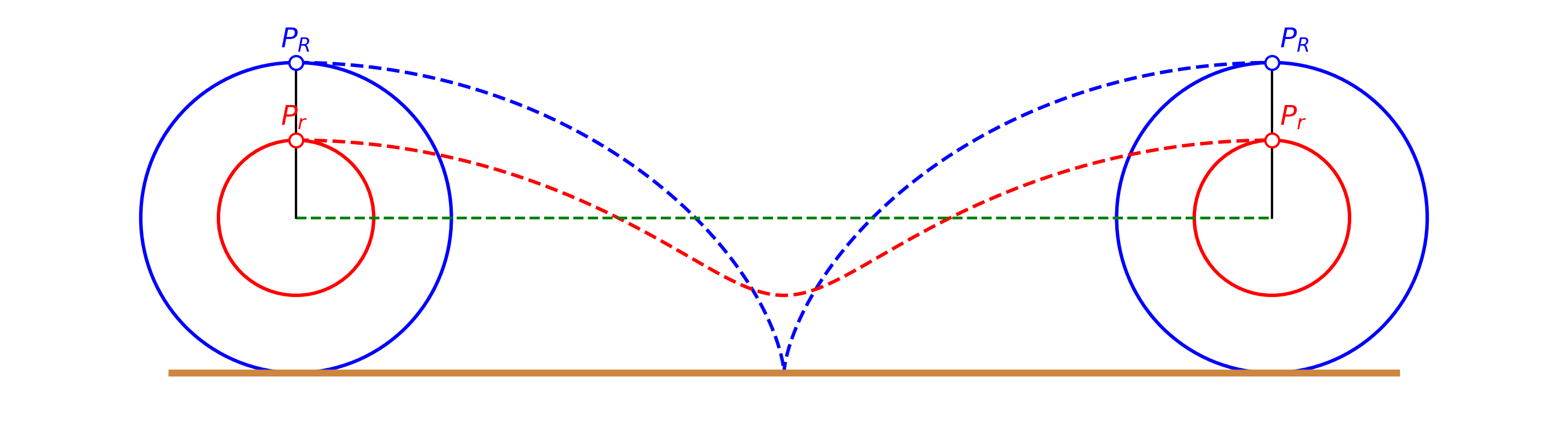}
    \caption{Illustration of the solution of Aristotle's wheel paradox for $k = 2$.}
    \label{fig9}
\end{figure}

Fig.~\ref{fig9} illustrates the wheel paradox: the circles before and after rolling one revolution, showing the motions of the center of point $P_R$ on the circle of radius $R$, and $P_r$ of the circle of radius $r$, with $P_R$ and $P_r$ starting and ending at the top of their respective circles. The green dashed line is the center's motion. The blue dashed curve shows $P_R$ motion. The red dash curve shows $P_r$ motion. Notice that the $P_r$ path is clearly shorter than the path of $P_R$. The ratio between the two circles in Fig.~\ref{fig9} is $k = R/r = 2$.

The closer $P_r$ is to the center, the shorter, more direct, and closer to the green line its path is. Fig.~\ref{fig10} below shows the ratio $k = 10$

\begin{figure}[h]
    \centering
    \includegraphics[width=0.9\textwidth]{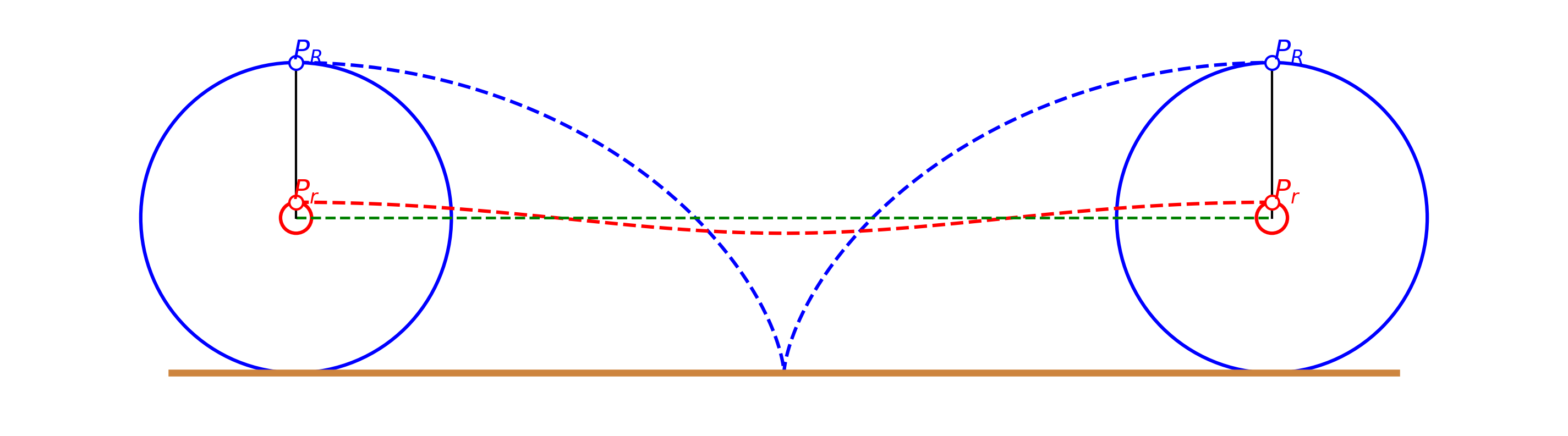}
    \caption{Illustration of the solution of Aristotle's wheel paradox for $k = 10$.}
    \label{fig10}
\end{figure}

The \emph{trochoids} \cite{Whitman1943, Yates1952, Wagon1999} is the name of the curve described by points $P_R$ in the bigger circle of radius $R$ and $P_r$ in the smaller circle of radius $r$, seen in Figs.\ref{fig9} and \ref{fig10} above. The trochoid is a curve formed when a circle rolls without slipping along a straight line. The kinematic constraint of rolling without slip imposes that the horizontal displacement $(x_c)$ of the circle’s center is proportional to the arc length traversed on its circumference, $x_c(\theta)=R\theta$, and the vertical coordinate of the center remains constant, $y_c(\theta) = R$ as the circle rolls over the line taken as the $x$-axis. If a point is located at a fixed distance $a$ from the center, its position in the laboratory frame results from the superposition of the translation of the circle’s center and the rotation of the point around the center. The parametric representation of its trajectory is therefore
\begin{align}
&x(\theta)= R\theta + a\sin\theta \\
&y(\theta)=R + a\cos\theta.
\end{align}
The geometrical aspects of the trochoid curve depend on the relation between the distance $a$ and the circle’s radius $R$. If $a=R$, the traced curve reduces to the \emph{cycloid}, the roulette generated by a circle rolling without slipping on a line. If $a<R$, the trajectory is a \emph{curtate trochoid}, characterized by arches that do not reach the baseline; while for $a>R$, one obtains a \emph{prolate trochoid}, whose lobes extend beyond the rolling line. The cycloid, as a particular instance, has played a fundamental role in the history of mathematics and physics, for example, the brachistochrone problem \cite{Courant1996, Thornton2004}.

\section{Python codes}

\subsection{Figure 1: circles with arrows}

To produce a proxy for Fig.~\ref{fig1}, but instead of coins, the figure shows circles with arrows, run the Python code provided below using a Jupyter Notebook.

\begin{lstlisting}[language=Python, numbers=none]
import matplotlib.pyplot as plt
from matplotlib.patches import Circle, FancyArrowPatch

def draw_equal_quincunx_with_arrows(radius=1.0, arrow_len=None, savepath=None):
    """
    Five equal circles (radius = radius) in quincunx.
    Outer four are tangent to the center circle (centers at distance d=2*radius).
    Dashed guide circle goes through the outer centers.
    Arrows start at the CENTER of each outer circle:
        top & bottom: upwards; left & right: downwards.
    The dashed guide is rendered BEHIND the disks (hidden where overlapped).
    """
    s = float(radius)
    if s <= 0:
        raise ValueError("radius must be positive.")

    d = 2.0 * s
    if arrow_len is None:
        arrow_len = 1.1 * s

    centers = {
        "top":    (0.0,  d),
        "right":  ( d,  0.0),
        "bottom": (0.0, -d),
        "left":   (-d,  0.0),
        "center": (0.0,  0.0),
    }

    # z-order: guide (1) < circles (3) < arrows (4)
    z_guide, z_circles, z_arrows = 1, 3, 4

    fig, ax = plt.subplots(figsize=(6, 6))
    ax.set_aspect('equal')
    ax.axis('off')

    # --- dashed guide circle FIRST (behind) ---
    ax.add_patch(Circle((0.0, 0.0), d, ec="#4a5568", fc="none",
                        lw=1.8, ls="--", zorder=z_guide))

    # --- disks on top of guide (opaque facecolors hide the dashed where overlapped) ---
    for key in ["top", "right", "bottom", "left", "center"]:
        cx, cy = centers[key]
        ax.add_patch(Circle((cx, cy), s, ec="#2b6cb0", fc="#e6eefc",
                            lw=2, zorder=z_circles))

    # --- arrows on top ---
    def vertical_arrow_from_center(cx, cy, direction):
        if direction == "up":
            start, end = (cx, cy), (cx, cy + arrow_len)
        elif direction == "down":
            start, end = (cx, cy), (cx, cy - arrow_len)
        else:
            raise ValueError("direction must be 'up' or 'down'")
        a = FancyArrowPatch(
            start, end, arrowstyle='-|>', mutation_scale=16,
            lw=2.0, color='#44515c', zorder=z_arrows
        )
        ax.add_patch(a)

    vertical_arrow_from_center(*centers["top"],    "up")
    vertical_arrow_from_center(*centers["bottom"], "up")
    vertical_arrow_from_center(*centers["left"],   "down")
    vertical_arrow_from_center(*centers["right"],  "down")

    # limits include arrows
    pad = 0.3 * s
    x_pad = s + pad
    y_pad = max(s, arrow_len) + pad
    ax.set_xlim(-d - x_pad, d + x_pad)
    ax.set_ylim(-d - y_pad, d + y_pad)

    if savepath:
        fig.savefig(savepath, dpi=300, bbox_inches='tight', pad_inches=0.1)
    plt.show()
    plt.close(fig)

# Example:
draw_equal_quincunx_with_arrows(radius=1.0, arrow_len=1.1)
\end{lstlisting}

\subsection{Figure 1: circles with arrows animated}

To produce a proxy for Fig.~\ref{fig1}, but with circles and arrows instead of coins, in animation style, run the Python code provided below in a Jupyter Notebook.

\begin{lstlisting}[language=Python, numbers=none]
import numpy as np
import matplotlib.pyplot as plt
from matplotlib.patches import Circle, FancyArrowPatch
from matplotlib.animation import FuncAnimation
from IPython.display import HTML

def animate_two_circles_with_roulette_and_arrow(radius=1.0, frames=360, interval=25,
                                                clockwise=False, savepath=None, fps=30, dpi=150):
    R = float(radius); r = R
    sgn = -1.0 if clockwise else 1.0
    phi0 = np.pi / 2
    phi  = phi0 + sgn * np.linspace(0.0, 2*np.pi, frames, endpoint=False)
    psi  = 2.0 * (phi - phi0)           # external rolling, R=r -> factor 2
    cx   = (R + r) * np.cos(phi)
    cy   = (R + r) * np.sin(phi)
    alpha0 = phi0 + np.pi               # marker initially toward origin
    Px = cx + r * np.cos(psi + alpha0)
    Py = cy + r * np.sin(psi + alpha0)

    fig, ax = plt.subplots(figsize=(6, 6))
    ax.set_aspect('equal'); ax.axis('off')
    center_disk  = Circle((0, 0), R, ec="#2b6cb0", fc="#e6eefc", lw=2, zorder=3)
    rolling_disk = Circle((cx[0], cy[0]), r, ec="#d65f5f", fc="#fdecec", lw=2, zorder=3)
    ax.add_patch(center_disk); ax.add_patch(rolling_disk)
    path_line, = ax.plot([], [], '-', lw=2.0, color='red', zorder=2)
    P_dot,     = ax.plot(Px[0], Py[0], 'o', ms=6, mfc='white', mec='#222', zorder=4)
    arrow = FancyArrowPatch((cx[0], cy[0]), (Px[0], Py[0]),
                            arrowstyle='-|>', mutation_scale=16, lw=2.0,
                            color='#44515c', zorder=4)
    ax.add_patch(arrow)

    pad = 0.4 * R
    ax.set_xlim(-3*R - pad, 3*R + pad)
    ax.set_ylim(-3*R - pad, 3*R + pad)

    def update(i):
        rolling_disk.center = (cx[i], cy[i])
        P_dot.set_data(Px[i], Py[i])
        path_line.set_data(Px[:i+1], Py[:i+1])
        ang = psi[i] + alpha0
        arr_len = 0.9 * R
        start = (cx[i], cy[i])
        end   = (cx[i] + arr_len*np.cos(ang), cy[i] + arr_len*np.sin(ang))
        arrow.set_positions(start, end)
        return rolling_disk, P_dot, path_line, arrow

    anim = FuncAnimation(fig, update, frames=frames, interval=interval, blit=True)

    # ---- Save if requested ----
    if savepath is not None:
        ext = savepath.split(".")[-1].lower()
        if ext in {"mp4", "m4v", "mov"}:
            from matplotlib.animation import FFMpegWriter
            anim.save(savepath, writer=FFMpegWriter(fps=fps), dpi=dpi)
        elif ext in {"gif"}:
            from matplotlib.animation import PillowWriter
            anim.save(savepath, writer=PillowWriter(fps=fps), dpi=dpi)
        else:
            raise ValueError("Unsupported extension; use .mp4 or .gif")

    plt.close(fig)   # close figure to avoid duplicate display in notebooks
    return HTML(anim.to_jshtml())

# Show inline only
animate_two_circles_with_roulette_and_arrow(radius=1.0)
# Save as MP4 (needs ffmpeg)
animate_two_circles_with_roulette_and_arrow(radius=1.0, savepath="roll.mp4", fps=30, dpi=150)
# Save as GIF (uses Pillow)
animate_two_circles_with_roulette_and_arrow(radius=1.0, savepath="roll.gif", fps=20)
\end{lstlisting}

\subsection{Figure 2: solving the paradox with physics}

To produce Fig.~\ref{fig2}, run the Python code provided below using a Jupyter Notebook, Google Colab Notebook, or your preferred development environment.

\begin{lstlisting}[language=Python, numbers=none]
import numpy as np
import matplotlib.pyplot as plt
from matplotlib.patches import Circle, FancyArrow, Arc

def draw_rolling_diagram(R=3, r=1, theta_deg=30, save_path=None):
    """
    Rolling-circle diagram with:
      - v drawn from the small-circle center (longer, tangent to center path),
      - w curved arrow placed on TOP of the small circle (outside) to avoid crossings,
      - R and r shown as teal radius segments with midpoint labels.
    """
    theta = np.deg2rad(theta_deg)
    RR = R + r
    O = (0.0, 0.0)
    C = (RR*np.cos(theta), RR*np.sin(theta))

    fig, ax = plt.subplots(figsize=(6, 6))
    col, teal = '#1f2d3a', '#1f6f7a'

    # Dashed path of the rolling center
    ax.add_patch(Circle(O, RR, fill=False, ls='--', lw=1.5, color=col))

    # Circles
    ax.add_patch(Circle(O, R, fill=False, lw=2, color=col))
    ax.add_patch(Circle(C, r, fill=False, lw=2, color=col))

    # Direction from O to C
    ang = np.arctan2(C[1]-O[1], C[0]-O[0])

    # Radius R with midpoint label
    R_end = (O[0] + R*np.cos(ang), O[1] + R*np.sin(ang))
    ax.plot([O[0], R_end[0]], [O[1], R_end[1]], color='#1f6f7a', lw=2)
    R_mid = (O[0] + 0.5*R*np.cos(ang), O[1] + 0.5*R*np.sin(ang))
    ax.text(R_mid[0], R_mid[1], r"$R$", color='#1f6f7a', fontsize=12, ha='center', va='bottom')

    # Radius r with midpoint label
    r_end = (C[0] - r*np.cos(ang), C[1] - r*np.sin(ang))
    ax.plot([C[0], r_end[0]], [C[1], r_end[1]], color='#1f6f7a', lw=2)
    r_mid = (C[0] - 0.5*r*np.cos(ang), C[1] - 0.5*r*np.sin(ang))
    ax.text(r_mid[0], r_mid[1], r"$r$", color='#1f6f7a', fontsize=12, ha='center', va='bottom')

    # Velocity vector from small-circle center (tangent)
    vx, vy = np.cos(ang + np.pi/2), np.sin(ang + np.pi/2)
    v_len = 1.8
    ax.add_patch(FancyArrow(C[0], C[1], v_len*vx, v_len*vy,
                            width=0.025, head_width=0.22, head_length=0.22,
                            color='#1f6f7a'))
    ax.text(C[0] + v_len*vx + 0.12, C[1] + v_len*vy, r"$\vec{v}$", color='#1f6f7a', fontsize=12)

    # w curved arrow OUTSIDE small circle, on top (absolute angles 70deg to 110 deg)
    arc_rad = 1.20 * r
    start_deg_abs, end_deg_abs = 70, 110
    arc = Arc(C, 2*arc_rad, 2*arc_rad, theta1=start_deg_abs, theta2=end_deg_abs, color='#1f6f7a', lw=2)
    ax.add_patch(arc)
    end_rad = np.deg2rad(end_deg_abs)
    hx = C[0] + arc_rad*np.cos(end_rad)
    hy = C[1] + arc_rad*np.sin(end_rad)
    tdx, tdy = -np.sin(end_rad), np.cos(end_rad)  # CCW tangent
    ax.add_patch(FancyArrow(hx - 0.12*tdx, hy - 0.12*tdy, 0.12*tdx, 0.12*tdy,
                            width=0.0, head_width=0.18, head_length=0.18,
                            color='#1f6f7a', length_includes_head=True))
    ax.text(hx + 0.22, hy + 0.18, r"$\omega$", color='#1f6f7a', fontsize=12)

    # Layout
    ax.set_aspect('equal', adjustable='box')
    ax.set_xlim(-RR-1.4, RR+1.8)
    ax.set_ylim(-RR-1.4, RR+1.8)
    ax.axis('off')

    if save_path:
        plt.savefig(save_path, dpi=300, bbox_inches='tight')
    return fig, ax

# Example usage:
fig, ax = draw_rolling_diagram(R=3, r=1, theta_deg=30, save_path="fig2.png")
\end{lstlisting}

\subsection{Figure 3: the epicycloid}

To produce Fig.~\ref{fig3}, one must execute the Python code provided below.

\begin{lstlisting}[language=Python, numbers=none]
import numpy as np
import matplotlib.pyplot as plt
from matplotlib.patches import Circle

def plot_epicycloid(R=3.0, r=1.0, points=2000, save_path=None, show=True):
    """
    Plot the epicycloid generated by a circle of radius r rolling
    around a fixed circle of radius R.

    Parameters
    ----------
    R : float
        Radius of the fixed circle.
    r : float
        Radius of the rolling circle.
    points : int
        Number of sample points for the epicycloid.
    save_path : str or None
        If provided, saves the figure to this path.
    show : bool
        If True, displays the figure. If False, closes it.
    
    Returns
    -------
    fig, ax : Matplotlib figure and axes objects
    """
    k = (R + r) / r
    theta = np.linspace(0, 2*np.pi, points)
    x = (R + r) * np.cos(theta) - r * np.cos(k * theta)
    y = (R + r) * np.sin(theta) - r * np.sin(k * theta)

    fig, ax = plt.subplots(figsize=(6,6))

    # Grid
    ax.set_xlim(-R-r-1, R+r+1)
    ax.set_ylim(-R-r-1, R+r+1)
    ax.set_aspect('equal', 'box')
    ax.set_xticks(np.arange(-R-r, R+r+1, 1))
    ax.set_yticks(np.arange(-R-r, R+r+1, 1))
    ax.grid(True, which='both', linestyle='-', linewidth=0.5, alpha=0.25)

    # Axes lines
    ax.axhline(0, color='k', linewidth=1)
    ax.axvline(0, color='k', linewidth=1)
    ax.text(R+r-0.5, -0.3, 'x', fontsize=12)
    ax.text(0.2, R+r-0.5, 'y', fontsize=12)

    # Fixed circle (blue)
    ax.add_patch(Circle((0,0), R, fill=False, edgecolor='blue', linewidth=2))

    # Rolling circle (black) at initial position
    rolling_center = (R + r, 0.0)
    ax.add_patch(Circle(rolling_center, r, fill=False, edgecolor='black', linewidth=2))

    # Center dot
    ax.plot(rolling_center[0], rolling_center[1], 'ko')

    # Epicycloid path
    ax.plot(x, y, color='red', linewidth=2)

    # Starting point
    ax.plot([R], [0], 'o', color='red')

    # Clean frame
    for spine in ax.spines.values():
        spine.set_visible(False)

    plt.tight_layout()

    if save_path:
        plt.savefig(save_path, dpi=300, bbox_inches='tight')

    if not show:
        plt.close(fig)

    return fig, ax

fig, ax = plot_epicycloid(R=3.0, r=1.0, save_path="fig3.png")
\end{lstlisting}

\subsection{Figure 3 animated: the epicycloid animation}

The Python code below produces Fig.~\ref{fig3} in animation style. Run the code in your Jupyter Notebook to see an MP4 video of the epicycloid being traced. 

\begin{lstlisting}[language=Python, numbers=none]
import numpy as np
import matplotlib.pyplot as plt
from matplotlib.patches import Circle
from matplotlib.animation import FuncAnimation
from matplotlib import rc

rc("animation", html="jshtml")   # for Jupyter animation display

def animate_epicycloid(R=3.0, r=1.0, frames=600, interval=20):
    """
    Animate the epicycloid generated by a circle of radius r rolling
    around a fixed circle of radius R.

    Parameters
    ----------
    R : float
        Radius of the fixed circle.
    r : float
        Radius of the rolling circle.
    frames : int
        Number of frames in the animation.
    interval : int
        Delay between frames in milliseconds.

    Returns
    -------
    ani : matplotlib.animation.FuncAnimation
        Animation object (renders inline in Jupyter).
    """
    k = (R + r) / r
    theta = np.linspace(0, 2*np.pi, frames)
    x = (R + r) * np.cos(theta) - r * np.cos(k * theta)
    y = (R + r) * np.sin(theta) - r * np.sin(k * theta)

    fig, ax = plt.subplots(figsize=(6,6))
    ax.set_xlim(-R-r-1, R+r+1)
    ax.set_ylim(-R-r-1, R+r+1)
    ax.set_aspect('equal', 'box')
    ax.set_xticks(np.arange(-R-r, R+r+1, 1))
    ax.set_yticks(np.arange(-R-r, R+r+1, 1))
    ax.grid(True, linestyle='-', linewidth=0.5, alpha=0.25)
    ax.axhline(0, color='k', linewidth=1)
    ax.axvline(0, color='k', linewidth=1)
    ax.text(R+r-0.5, -0.3, 'x')
    ax.text(0.2, R+r-0.5, 'y')
    for s in ax.spines.values(): 
        s.set_visible(False)

    # Fixed circle
    ax.add_patch(Circle((0,0), R, fill=False, edgecolor='blue', linewidth=2))

    # Rolling circle
    rolling_circle = Circle((R+r, 0), r, fill=False, edgecolor='black', linewidth=2)
    ax.add_patch(rolling_circle)

    # Traced point and path
    trace_dot, = ax.plot([], [], 'ro')
    trace_line, = ax.plot([], [], 'r-', linewidth=2)

    # Starting point
    ax.plot([R], [0], 'o', color='red')
    plt.close(fig)

    def init():
        rolling_circle.center = (R+r, 0)
        trace_dot.set_data([], [])
        trace_line.set_data([], [])
        return trace_dot, trace_line, rolling_circle

    def update(i):
        cx = (R + r) * np.cos(theta[i])
        cy = (R + r) * np.sin(theta[i])
        rolling_circle.center = (cx, cy)
        trace_dot.set_data(x[i], y[i])
        trace_line.set_data(x[:i+1], y[:i+1])
        return trace_dot, trace_line, rolling_circle

    ani = FuncAnimation(fig, update, init_func=init,
                        frames=frames, interval=interval, blit=True)
    return ani
ani = animate_epicycloid(R=3.0, r=1.0, frames=600, interval=30)
ani  # shows animation
\end{lstlisting}

\subsection{Figure 5: several examples of epicycloids}

The Python code below produces Fig.~\ref{fig5}. Run the code in your Jupyter Notebook or Google Colab Notebook.

\begin{lstlisting}[language=Python, numbers=none]
import numpy as np
import matplotlib.pyplot as plt

def epicycloid_xy(k, num_points=2400, turns=1.0):
    r = 1.0
    R = k * r
    w = (R + r) / r
    t = np.linspace(0, 2 * np.pi * turns, num_points)
    x = (R + r) * np.cos(t) - r * np.cos(w * t)
    y = (R + r) * np.sin(t) - r * np.sin(w * t)
    return x, y

def plot_panel(ax, k, subtitle=None):
    turns = 1.0 if abs(k - round(k)) < 1e-9 else 12.0
    x, y = epicycloid_xy(k, num_points=4000, turns=turns)
    ax.axhline(0, color='0.6', lw=0.8, ls='--')
    ax.axvline(0, color='0.6', lw=0.8, ls='--')
    ax.plot(x, y, color='black', lw=1.8)
    ax.set_aspect('equal', 'box')
    lim = k + 3.0
    ax.set_xlim(-lim, lim)
    ax.set_ylim(-lim, lim)
    ax.set_xticks([]); ax.set_yticks([])
    for spine in ax.spines.values():
        spine.set_linewidth(1.0)
        spine.set_color('0.3')
    if subtitle:
        ax.text(0.5, -0.10, subtitle, transform=ax.transAxes,
                ha='center', va='top', fontsize=22)

# Parameters
k_list = [1, 2, 3, 4, 2.5, 5.8, 25, 100]
subs = ["k = 1; cardioid",
        "k = 2; nephroid",
        "k = 3; trefoiloid",
        "k = 4; quatrefoiloid",
        "k = 2.5",
        "k = 5.8",
        "k = 25",
        "k = 100"]

# Make subplot areas bigger
fig, axes = plt.subplots(2, 4, figsize=(22, 12))  # Bigger figure = bigger boxes
axes = axes.ravel()

for i in range(len(k_list)):
    plot_panel(axes[i], k_list[i], subtitle=subs[i])

# Make panels occupy more space
plt.subplots_adjust(left=0.00, right=1, top=1, bottom=0.1,
                    wspace=0.1, hspace=0.3)  # Smaller wspace/hspace = bigger boxes

# Show or save
plt.savefig("fig5.png", dpi=300, bbox_inches="tight")
plt.show()
\end{lstlisting}

\subsection{Figure 6: two different views of the coin paradox}

To produce Fig.~\ref{fig6} run the Python code bellow

\begin{lstlisting}[language=Python, numbers=none]
import matplotlib.pyplot as plt
from matplotlib.patches import Circle, FancyArrow
import numpy as np

def draw_coin(ax, center, radius, facecolor='#87CEEB', edgecolor='black',
              arm_angle=None, arm_len=None, armcolor='black', lw=3):
    """Draw a coin with an optional orientation arrow."""
    circ = Circle(center, radius, facecolor=facecolor, edgecolor=edgecolor, lw=lw)
    ax.add_patch(circ)
    # center dot
    ax.plot(center[0], center[1], 'o', color='black', ms=4)
    # orientation arrow
    if arm_angle is not None:
        if arm_len is None:
            arm_len = radius * 0.9
        x0, y0 = center
        x1 = x0 + arm_len * np.cos(arm_angle)
        y1 = y0 + arm_len * np.sin(arm_angle)
        arr = FancyArrow(x0, y0, x1 - x0, y1 - y0,
                         width=0.03*radius, head_width=0.25*radius, head_length=0.25*radius,
                         color=armcolor, length_includes_head=True)
        ax.add_patch(arr)

def tangent_center(big_c, R, r, angle):
    """Center of a small circle tangent to the big circle at polar angle 'angle'."""
    return (big_c[0] + (R + r) * np.cos(angle),
            big_c[1] + (R + r) * np.sin(angle))

def draw_scene(ax, big_center, R, r, mode='left',
               big_color='#87CEEB', small_color='#B0E0E6'):
    """Draw one panel (left or right)."""
    # Big coin
    draw_coin(ax, big_center, R, facecolor=big_color, lw=3)

    # Top small coin (tangent at 90) with downward arrow
    top_c = tangent_center(big_center, R, r, np.pi/2)
    draw_coin(ax, top_c, r, facecolor=big_color, lw=3, arm_angle=-np.pi/2)

    if mode == 'left':
        # Two tangent small coins at 210 and 330 with angled arrows
        c1 = tangent_center(big_center, R, r, np.deg2rad(210))
        c2 = tangent_center(big_center, R, r, np.deg2rad(330))
        draw_coin(ax, c1, r, facecolor=small_color, edgecolor='#6b6b6b',
                  arm_angle=np.deg2rad(30))
        draw_coin(ax, c2, r, facecolor=small_color, edgecolor='#6b6b6b',
                  arm_angle=np.deg2rad(150))
    else:
        # Three tangent coins: left (180), right (0), bottom (270)
        left_c   = tangent_center(big_center, R, r, np.pi)
        right_c  = tangent_center(big_center, R, r, 0.0)
        bottom_c = tangent_center(big_center, R, r, -np.pi/2)

        # Left & right coins: arrows up
        draw_coin(ax, left_c,  r, facecolor=small_color, edgecolor='#6b6b6b',
                  arm_angle=-np.pi/2)
        draw_coin(ax, right_c, r, facecolor=small_color, edgecolor='#6b6b6b',
                  arm_angle=-np.pi/2)

        # Bottom coin: arrow DOWN (fixed)
        draw_coin(ax, bottom_c, r, facecolor=small_color, edgecolor='#6b6b6b',
                  arm_angle=-np.pi/2)

def print_image(r=0.6, k=3, center_distance=8.0, save_path=None):
    """Render the two panels and optionally save to file."""
    R = k * r  # enforce R = k r

    fig, ax = plt.subplots(figsize=(10, 5))
    left_center  = (-center_distance/2, 0.0)
    right_center = ( center_distance/2, 0.0)

    draw_scene(ax, left_center,  R, r, mode='left')
    draw_scene(ax, right_center, R, r, mode='right')

    # Labels (a) and (b) under each panel
    y_label = - (R + r) - 0.8
    ax.text(left_center[0],  y_label, '(a)', fontsize=14, ha='center', va='top')
    ax.text(right_center[0], y_label, '(b)', fontsize=14, ha='center', va='top')

    # Layout
    extent = center_distance/2 + (R + r) + 1.0
    ax.set_xlim(-extent, extent)
    ax.set_ylim(- (R + r) - 1.5, (R + r) + 1.5)
    ax.set_aspect('equal', adjustable='box')
    ax.axis('off')

    if save_path:
        plt.savefig(save_path, dpi=300, bbox_inches='tight')
    plt.show()

# Example
print_image(r=0.6, k=3, center_distance=8.0, save_path="fig6.png")
\end{lstlisting}

\subsection{Figure 7: the hypocycloid}

To produce Fig.~\ref{fig7} run the Python code bellow

\begin{lstlisting}[language=Python, numbers=none]
import numpy as np
import matplotlib.pyplot as plt
from matplotlib.patches import Circle

def plot_epicycloid(R=3.0, r=1.0, points=2000, save_path=None, show=True):
    """
    Plot the epicycloid generated by a circle of radius r rolling
    around a fixed circle of radius R.

    Parameters
    ----------
    R : float
        Radius of the fixed circle.
    r : float
        Radius of the rolling circle.
    points : int
        Number of sample points for the epicycloid.
    save_path : str or None
        If provided, saves the figure to this path.
    show : bool
        If True, displays the figure. If False, closes it.
    
    Returns
    -------
    fig, ax : Matplotlib figure and axes objects
    """
    k = - (R - r) / r
    theta = np.linspace(0, 2*np.pi, points)
    x = (R - r) * np.cos(theta) + r * np.cos(k * theta)
    y = (R - r) * np.sin(theta) + r * np.sin(k * theta)

    fig, ax = plt.subplots(figsize=(6,6))

    # Grid
    ax.set_xlim(-R-r-1, R+r+1)
    ax.set_ylim(-R-r-1, R+r+1)
    ax.set_aspect('equal', 'box')
    ax.set_xticks(np.arange(-R-r, R+r+1, 1))
    ax.set_yticks(np.arange(-R-r, R+r+1, 1))
    ax.grid(True, which='both', linestyle='-', linewidth=0.5, alpha=0.25)

    # Axes lines
    ax.axhline(0, color='k', linewidth=1)
    ax.axvline(0, color='k', linewidth=1)
    ax.text(R+r-0.5, -0.3, 'x', fontsize=12)
    ax.text(0.2, R+r-0.5, 'y', fontsize=12)

    # Fixed circle (blue)
    ax.add_patch(Circle((0,0), R, fill=False, edgecolor='blue', linewidth=2))

    # Rolling circle (black) at initial position
    rolling_center = (R - r, 0.0)
    ax.add_patch(Circle(rolling_center, r, fill=False, edgecolor='black', linewidth=2))

    # Center dot
    ax.plot(rolling_center[0], rolling_center[1], 'ko')

    # Epicycloid path
    ax.plot(x, y, color='red', linewidth=2)

    # Starting point
    ax.plot([R], [0], 'o', color='red')

    # Clean frame
    for spine in ax.spines.values():
        spine.set_visible(False)

    plt.tight_layout()

    if save_path:
        plt.savefig(save_path, dpi=300, bbox_inches='tight')

    if not show:
        plt.close(fig)

    return fig, ax

fig, ax = plot_epicycloid(R=3.0, r=1.0, save_path="fig7.png")
\end{lstlisting}

\subsection{Figure 7: the animated hypocycloid}

The Python code below produces Fig.~\ref{fig7} in animation style. Run the code in your Jupyter Notebook to see an MP4 video of the epicycloid being traced. 

\begin{lstlisting}[language=Python, numbers=none]
import numpy as np
import matplotlib.pyplot as plt
from matplotlib.patches import Circle
from matplotlib.animation import FuncAnimation
from matplotlib import rc

rc("animation", html="jshtml")   # for Jupyter animation display

def animate_epicycloid(R=3.0, r=1.0, frames=600, interval=20):
    """
    Animate the epicycloid generated by a circle of radius r rolling
    around a fixed circle of radius R.

    Parameters
    ----------
    R : float
        Radius of the fixed circle.
    r : float
        Radius of the rolling circle.
    frames : int
        Number of frames in the animation.
    interval : int
        Delay between frames in milliseconds.

    Returns
    -------
    ani : matplotlib.animation.FuncAnimation
        Animation object (renders inline in Jupyter).
    """
    k = - (R - r) / r
    theta = np.linspace(0, 2*np.pi, frames)
    x = (R - r) * np.cos(theta) + r * np.cos(k * theta)
    y = (R - r) * np.sin(theta) + r * np.sin(k * theta)

    fig, ax = plt.subplots(figsize=(6,6))
    ax.set_xlim(-R-r-1, R+r+1)
    ax.set_ylim(-R-r-1, R+r+1)
    ax.set_aspect('equal', 'box')
    ax.set_xticks(np.arange(-R-r, R+r+1, 1))
    ax.set_yticks(np.arange(-R-r, R+r+1, 1))
    ax.grid(True, linestyle='-', linewidth=0.5, alpha=0.25)
    ax.axhline(0, color='k', linewidth=1)
    ax.axvline(0, color='k', linewidth=1)
    ax.text(R+r-0.5, -0.3, 'x')
    ax.text(0.2, R+r-0.5, 'y')
    for s in ax.spines.values(): 
        s.set_visible(False)

    # Fixed circle
    ax.add_patch(Circle((0,0), R, fill=False, edgecolor='blue', linewidth=2))

    # Rolling circle
    rolling_circle = Circle((R+r, 0), r, fill=False, edgecolor='black', linewidth=2)
    ax.add_patch(rolling_circle)

    # Traced point and path
    trace_dot, = ax.plot([], [], 'ro')
    trace_line, = ax.plot([], [], 'r-', linewidth=2)

    # Starting point
    ax.plot([R], [0], 'o', color='red')
    plt.close(fig)

    def init():
        rolling_circle.center = (R+r, 0)
        trace_dot.set_data([], [])
        trace_line.set_data([], [])
        return trace_dot, trace_line, rolling_circle

    def update(i):
        cx = (R - r) * np.cos(theta[i])
        cy = (R - r) * np.sin(theta[i])
        rolling_circle.center = (cx, cy)
        trace_dot.set_data(x[i], y[i])
        trace_line.set_data(x[:i+1], y[:i+1])
        return trace_dot, trace_line, rolling_circle

    ani = FuncAnimation(fig, update, init_func=init,
                        frames=frames, interval=interval, blit=True)
    return ani

ani = animate_epicycloid(R=3.0, r=1.0, frames=600, interval=30)
ani  # shows animation
\end{lstlisting}

\subsection{Figure 8: several examples of hypocycloid}

To produce Fig.~\ref{fig8} run the Python code bellow

\begin{lstlisting}[language=Python, numbers=none]
import numpy as np
import matplotlib.pyplot as plt

def epicycloid_xy(k, num_points=2400, turns=1.0):
    r = 1.0
    R = k * r
    w = - (R - r) / r
    t = np.linspace(0, 2 * np.pi * turns, num_points)
    x = (R - r) * np.cos(t) + r * np.cos(w * t)
    y = (R - r) * np.sin(t) + r * np.sin(w * t)
    return x, y

def plot_panel(ax, k, subtitle=None):
    turns = 1.0 if abs(k - round(k)) < 1e-9 else 12.0
    x, y = epicycloid_xy(k, num_points=4000, turns=turns)
    ax.axhline(0, color='0.6', lw=0.8, ls='--')
    ax.axvline(0, color='0.6', lw=0.8, ls='--')
    ax.plot(x, y, color='black', lw=1.8)
    ax.set_aspect('equal', 'box')
    lim = k + 3.0
    ax.set_xlim(-lim, lim)
    ax.set_ylim(-lim, lim)
    ax.set_xticks([]); ax.set_yticks([])
    for spine in ax.spines.values():
        spine.set_linewidth(1.0)
        spine.set_color('0.3')
    if subtitle:
        ax.text(0.5, -0.10, subtitle, transform=ax.transAxes,
                ha='center', va='top', fontsize=22)

# Parameters
k_list = [3, 4, 5, 6, 2.1, 3.8, 5.5, 100]
subs = ["k = 3; deltoid",
        "k = 4; astroid",
        "k = 5",
        "k = 6",
        "k = 2.1",
        "k = 3.8",
        "k = 5.5",
        "k = 100"]

# Make subplot areas bigger
fig, axes = plt.subplots(2, 4, figsize=(22, 12))  # Bigger figure = bigger boxes
axes = axes.ravel()

for i in range(len(k_list)):
    plot_panel(axes[i], k_list[i], subtitle=subs[i])

# Make panels occupy more space
plt.subplots_adjust(left=0.00, right=1, top=1, bottom=0.1,
                    wspace=0.1, hspace=0.3)  # Smaller wspace/hspace = bigger boxes

# Show or save
plt.savefig("fig8.png", dpi=300, bbox_inches="tight")
plt.show()
\end{lstlisting}

\subsection{Figure 9: Aristotle's wheel paradox}

To produce Fig.~\ref{fig9} run the Python code bellow

\begin{lstlisting}[language=Python, numbers=none]
import numpy as np
import matplotlib.pyplot as plt

def plot_aristotle_wheel(R=2.0, r=1.0, savepath=None):
    """
    Plot Aristotle's Wheel Paradox diagram for two concentric rolling circles.

    Parameters
    ----------
    R : float
        Radius of the larger circle.
    r : float
        Radius of the smaller circle.
    savepath : str or None
        If provided, saves the figure to the given path.
    """
    theta = np.linspace(0, 2 * np.pi, 600)

    # General trochoid for a point at distance a from center
    def trochoid(a):
        x = R * theta + a * np.sin(theta)
        y = R + a * np.cos(theta)
        return x, y

    # Trajectories for P_R (a=R) and P_r (a=r)
    x_R, y_R = trochoid(R)
    x_r, y_r = trochoid(r)

    # Center line at y = R, spanning exactly 2 pi R
    x_line = [0.0, 2 * np.pi * R]
    y_line = [R, R]

    # Figure/axes
    fig, ax = plt.subplots(figsize=(12, 4))
    ax.set_aspect('equal')
    ax.axis('off')

    # Paths
    ax.plot(x_R, y_R, 'b--', lw=1.5, label="Path of $P_R$")
    ax.plot(x_r, y_r, 'r--', lw=1.5, label="Path of $P_r$")
    ax.plot(x_line, y_line, 'g--', lw=1.2, label="Center line")

    # Vertical black lines at start and end (center to top of large circle)
    ax.plot([0, 0], [R, R + R], 'k', lw=1)
    ax.plot([2 * np.pi * R, 2 * np.pi * R], [R, R + R], 'k', lw=1)

    # Circles at start (x=0) and end (x=2 pi R)
    left_large  = plt.Circle((0.0, R), R, edgecolor='blue', fill=False, lw=1.5)
    left_small  = plt.Circle((0.0, R), r, edgecolor='red',  fill=False, lw=1.5)
    right_large = plt.Circle((2 * np.pi * R, R), R, edgecolor='blue', fill=False, lw=1.5)
    right_small = plt.Circle((2 * np.pi * R, R), r, edgecolor='red',  fill=False, lw=1.5)
    for c in (left_large, left_small, right_large, right_small):
        ax.add_patch(c)

    # Markers and labels: top points on each circle at start/end
    PR_left  = (0.0, R + R)
    Pr_left  = (0.0, R + r)
    PR_right = (2 * np.pi * R, R + R)
    Pr_right = (2 * np.pi * R, R + r)

    ax.plot(*PR_left,  'bo', markersize=6, mfc='white')
    ax.plot(*Pr_left,  'ro', markersize=6, mfc='white')
    ax.plot(*PR_right, 'bo', markersize=6, mfc='white')
    ax.plot(*Pr_right, 'ro', markersize=6, mfc='white')

    ax.text(PR_left[0]  - 0.2, PR_left[1]  + 0.2, r"$P_R$", fontsize=12, color='blue')
    ax.text(Pr_left[0]  - 0.2, Pr_left[1]  + 0.2, r"$P_r$", fontsize=12, color='red')
    ax.text(PR_right[0] + 0.1, PR_right[1] + 0.2, r"$P_R$", fontsize=12, color='blue')
    ax.text(Pr_right[0] + 0.1, Pr_right[1] + 0.2, r"$P_r$", fontsize=12, color='red')

    # Ground line
    x_pad = max(R, r) * 0.8
    ax.plot([-x_pad, 2 * np.pi * R + x_pad], [0, 0], color='peru', lw=3)

    # Robust limits: include full circles and labels for any R, r
    pad_y = max(R, r) * 0.3
    y_min = min(0.0, R - max(R, r)) - pad_y
    y_max = R + max(R, r) + pad_y
    x_min = -R - x_pad
    x_max = 2 * np.pi * R + R + x_pad

    ax.set_xlim(x_min, x_max)
    ax.set_ylim(y_min, y_max)

    if savepath:
        fig.savefig(savepath, dpi=300, bbox_inches='tight', pad_inches=0.1)

    plt.close(fig)  # close to avoid handle leaks
    return fig

# Save as PNG
fig = plot_aristotle_wheel(R=2, r=1, savepath="fig9.png")  
fig
\end{lstlisting}

\subsection{Figure 9: Aristotle's wheel paradox animated}

To produce Fig.~\ref{fig9} in animation style, run the Python code below

\begin{lstlisting}[language=Python, numbers=none]
import numpy as np
import matplotlib.pyplot as plt
from matplotlib.patches import Circle
from matplotlib.animation import FuncAnimation
from IPython.display import HTML

def animate_aristotle_wheel(R=2.0, r=1.0, frames=300, interval=30, show_trails=True):
    """
    Animate Aristotle's Wheel Paradox: two concentric circles of radii R and r
    roll without slipping over a distance 2piR. Points P_R (a=R) and P_r (a=r)
    trace trochoids. Returns HTML for inline display in Jupyter.
    """
    # Parameter over one full revolution
    theta = np.linspace(0.0, 2*np.pi, frames)

    # Helpers
    def trochoid(a, th):
        x = R*th + a*np.sin(th)
        y = R   + a*np.cos(th)
        return x, y

    # Precompute paths for trails
    xR_path, yR_path = trochoid(R, theta)   # P_R path
    xr_path, yr_path = trochoid(r, theta)   # P_r path

    # Figure and axes
    fig, ax = plt.subplots(figsize=(12, 4))
    ax.set_aspect('equal')
    ax.axis('off')

    # Static floor and center line
    x_total = 2*np.pi*R
    pad_x = max(R, r)*0.8
    ax.plot([-pad_x, x_total + pad_x], [0, 0], color='peru', lw=3)   # floor
    ax.plot([0, x_total], [R, R], 'g--', lw=1.2)                     # center line y=R

    # Static end markers: vertical lines at start and end
    ax.plot([0, 0], [R, R+R], 'k', lw=1)
    ax.plot([x_total, x_total], [R, R+R], 'k', lw=1)

    # Dynamic patches: rolling concentric circles (share same center)
    big = Circle((0, R), R, fill=False, lw=1.8, ec='blue')
    small = Circle((0, R), r, fill=False, lw=1.8, ec='red')
    ax.add_patch(big)
    ax.add_patch(small)

    # Moving points P_R and P_r
    PR_dot, = ax.plot([], [], 'o', ms=6, mfc='white', mec='blue')
    Pr_dot, = ax.plot([], [], 'o', ms=6, mfc='white', mec='red')

    # Labels near points
    PR_lbl = ax.text(0, 0, r"$P_R$", fontsize=12, color='blue',
                     ha='left', va='bottom')
    Pr_lbl = ax.text(0, 0, r"$P_r$", fontsize=12, color='red',
                     ha='left', va='bottom')

    # Optional trails
    PR_trail, = ax.plot([], [], 'b--', lw=1.2) if show_trails else (None,)
    Pr_trail, = ax.plot([], [], 'r--', lw=1.2) if show_trails else (None,)

    # Robust limits (avoid clipping for any R, r)
    pad_y = max(R, r)*0.35
    y_min = min(0.0, R - max(R, r)) - pad_y
    y_max = R + max(R, r) + pad_y
    x_min = -R - pad_x
    x_max = x_total + R + pad_x
    ax.set_xlim(x_min, x_max)
    ax.set_ylim(y_min, y_max)

    # Animation update
    def update(i):
        th = theta[i]
        # Rolling center (translation only; rotation is encoded in trochoid points)
        center_x = R*th
        center_y = R
        big.center = (center_x, center_y)
        small.center = (center_x, center_y)

        # Current P_R and P_r
        xR, yR = trochoid(R, th)
        xr, yr = trochoid(r, th)

        PR_dot.set_data(xR, yR)
        Pr_dot.set_data(xr, yr)

        PR_lbl.set_position((xR + 0.08*R, yR + 0.08*R))
        Pr_lbl.set_position((xr + 0.08*R, yr + 0.08*R))

        # Update trails
        artists = [big, small, PR_dot, Pr_dot, PR_lbl, Pr_lbl]
        if show_trails:
            PR_trail.set_data(xR_path[:i+1], yR_path[:i+1])
            Pr_trail.set_data(xr_path[:i+1], yr_path[:i+1])
            artists += [PR_trail, Pr_trail]
        return artists

    anim = FuncAnimation(fig, update, frames=frames, interval=interval, blit=True)
    plt.close(fig)  # prevent duplicate static output
    return anim

# Example (in a Jupyter cell):
anim = animate_aristotle_wheel(R=2.0, r=1.0)
# Save to MP4 (requires ffmpeg installed)
anim.save("aristotle_wheel.mp4", writer="ffmpeg", dpi=150)
# Or save to GIF (requires ImageMagick or Pillow support)
anim.save("aristotle_wheel.gif", writer="pillow", dpi=100)
# Display animation
HTML(anim.to_jshtml())
\end{lstlisting}

\section{Conclusion}

This work demonstrated that the coin paradox, Aristotle’s wheel paradox, and the geometry of epicycloids and hypocycloids follow directly from the no-slip rolling constraint, which yields the rotation count $N=(R\pm r)/r$ for a circle of radius $r$ rolling on or inside a circle of radius $R$. The direct application of basic concepts in circular kinematics was used to solve apparent paradoxes and establish links between classical mechanics, geometry, and visualization.  

Parametric equations for epicycloids and hypocycloids were derived and illustrated using Python-generated figures, providing a computational toolset for teaching kinematics and curve geometry. Because roulette connects rotation, constraint motion, and analytic geometry with minimal prerequisites, it offers a compact, visually compelling framework for undergraduate instruction in physics and mathematics. As a bonus, Python code for animated figures was included.

The use of visual aids, such as computer-generated animations and interactive computational tools, can significantly enhance the pedagogy of physics education. Platforms such as YouTube, with channels like Veritasium \cite{VeritasiumSAT2023} and 3Blue1Brown \cite{3Blue1Brown}, have become commonplace resources for undergraduate and graduate students seeking to complement traditional university instruction with accessible, visually engaging explanations. 

All the codes that generated the images, and some more material with some animated figures can be found in a GitHub repo, in Ref. \cite{github}

\section{Data Availability}

No new data were created or analyzed in this study.

\end{document}